\documentclass[showpacs,amssymb,amsmath,nobibnotes, aps,prl,secnumarabic
,twocolumn%
]{revtex4}
\pdfoutput=1
\usepackage{bm}
\usepackage{natbib}
\usepackage{graphicx}

\begin{document}
\textfloatsep 10pt

\title{Capillary rogue waves}

\author{M. Shats}
\email{Michael.Shats@anu.edu.au}
\author{H. Punzmann}
\author{H. Xia}

\affiliation{Research School of
Physics and Engineering, The Australian National
University, Canberra ACT 0200, Australia}

\date{\today}

\begin{abstract}
We report the first observation of extreme wave events (rogue waves) in parametrically driven capillary waves. Rogue waves are observed above a certain threshold in forcing. Above this threshold, frequency spectra broaden and develop exponential tails. For the first time we present evidence of strong four-wave coupling in non-linear waves (high tricoherence), which points to modulation instability as the main mechanism in rogue waves. The generation of rogue waves is identified as the onset of a distinct tail in the probability density function of the wave heights. Their probability is higher than expected from the measured wave background.
\end{abstract}


\maketitle

Extreme wave events, also referred to as rogue or freak waves, are mostly known as an oceanic phenomenon responsible for a large number of maritime disasters. These waves, which have heights and steepness much greater than expected from the sea state \cite{Dysthe2008}, have become a focus of intense research in the last decade. Recently, large wave events were observed in other wave systems: optical rogue waves \cite{Solli2007} and acoustic waves in superfluid helium \cite{Ganshin2008}. These discoveries indicate that rogue waves may be rather universal. 

The rarity of these events in the ocean and obvious difficulties in their systematic characterization restrict the development of theoretical models of the rogue wave generation. Currently four to five competing hypotheses of rogue waves are considered \cite{Slunyaev2009}. Among them is the nonlinear mechanism of the wave-wave interactions, such as modulation instability. However, no direct evidence in support of this hypothesis has yet been found. 

Further progress in the rogue wave studies can be made by reinforcing that they constitute a new class of wave phenomena observed in different physics contexts. An ultimate goal of the research into the physics of rogue waves is to establish mechanisms responsible for their generation in order to (a) reliably predict their probability (in engineering applications), (b) generate and control such waves (e.g. in optical fibers), and (c) to avoid or suppress them (in the ocean). 

Among the key questions related to the studies of rogue waves are: (1) does the probability of rogue waves depend on the surrounding wave amplitudes, and (2) are there measurable characteristics of the wave background which would be indicative of the rogue wave probability? Modulation instability (see \cite{Zakharov2009} and references therein) has long been considered a likely mechanism for the generation of rogue waves owing to its universality. It has been found in surface gravity waves (Benjamin-Feir instability \cite{Benjamin1967}), Rossby and drift waves (\cite{Smolyakov2000,Connaughton2009} and references therein), Langmuir waves in plasma, and optical waves \cite{Zakharov2009}. Modulation instability is considered an important factor affecting the probability density function of the surface elevation in surface gravity waves \cite{Onorato2009}. It is also a major factor in the generation of optical rogue waves due, for example, to collisions of breathers found in numerical simulations of nonlinear Schr\"{o}dinger equation \cite{Akhmediev2009}. Though on the conceptual level there is an agreement on the importance of modulation instability for the rogue wave generation, there is no direct experimental evidence of the relation between the instability and probability of rogue waves.  

In this Letter we report the first observation of extreme wave events in nonlinear capillary waves. We show that the probability of capillary rogue waves strongly increases in a nonlinear stage of modulation instability. We present the first computation of the tricoherence which suggests that the degree of coherent four-wave coupling at this stage is high.  

\begin{figure}
\includegraphics[width=6.0 cm]{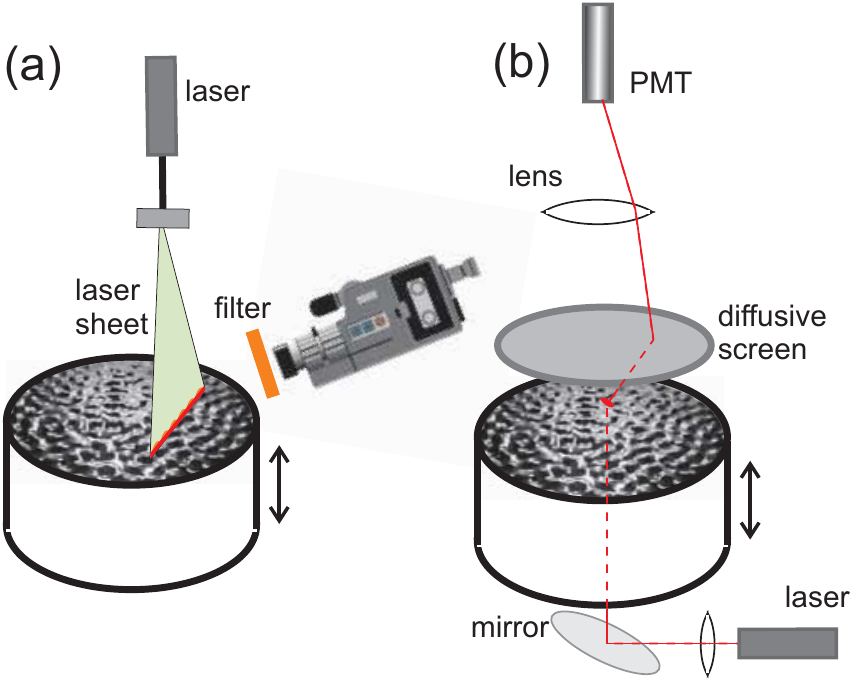}
\caption{\label{fig1} Experimental setup: (a) Fast video recording of the fluorescing water surface, (b) local measurement of the laser beam transmission through diffusing liquid.}
\end{figure}

Capillary waves belong to the higher-frequency branch of the surface waves, for which the restoring force is the surface tension. Their wavelengths are shorter than about 10 mm. Due to their smaller scale, capillary waves can be studied under well-controlled conditions in the laboratory using a variety of experimental methods to describe their space-time statistics and nonlinear wave-wave interactions. 
Capillary waves in our experiments are excited parametrically in a vertically shaken container filled with water whose depth is much greater than the wavelength \cite{Punzmann2009}. In these experiments waves are forced by shaking the container at the frequency of 60 Hz with the acceleration in the range of $a$ = (0.3 - 5)$g$. It should be noted that the results reported here are not sensitive to the choice of forcing frequency. The strength of forcing is characterised by the value of supercriticality above the threshold of parametric excitation $a_{th}$, $\varepsilon = a/a_{th}-1$. The strongest parametrically excited wave is the first subharmonic of the forcing frequency at $f_0$ = 30 Hz and several of its harmonics of smaller amplitudes. We developed several techniques to capture the dynamics of nonlinear capillary waves in a broad range of amplitudes with high temporal and spatial resolution. In the most direct method the perturbed surface is visualized by adding a small amount of a fluorescent dye (Rhodamine B) to the water and by illuminating it using a thin (0.2 mm x 16 mm) green laser sheet covering almost two wavelengths at the main frequency. The motion of the (orange) fluorescent surface in the vertical cross-section is then captured by fast video camera at 600 frames per second, Fig.~\ref{fig1}(a). An orange optical filter in front of the camera lens discriminates against the direct reflection of green laser light into the camera. The surface contour is detected using standard edge detection techniques to produce time-resolved records of the surface elevation. This technique works reliably at high wave amplitudes for waves of arbitrary steepness. 

\begin{figure}
\includegraphics[width=6.5 cm]{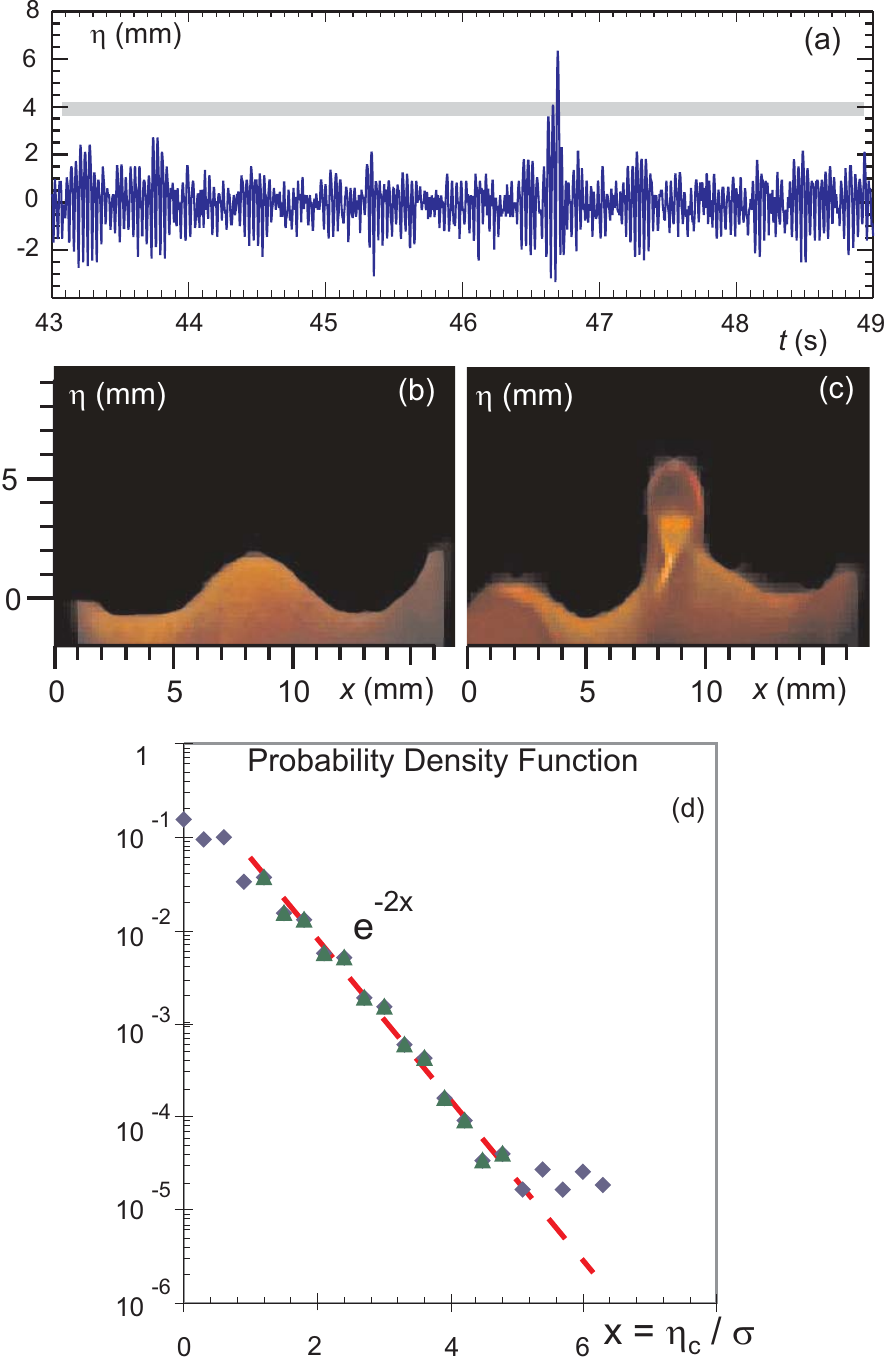}
\caption{\label{fig2} Phenomenology of capillary rogue waves at high level of forcing, $\varepsilon$ = 5. (a) Time trace of the surface elevation showing an extreme wave event in parametrically driven capillary waves. Video frames show waveforms (b) four wave periods before the large event, and (c) during the large wave event (gradient of the light intensity inside steep waves is due to the light refraction). (d) Probability density function (PDF) of the wave crests versus normalized crest height.}
\end{figure}

Figure~\ref{fig2} shows a time trace of the surface elevation $\eta(t)$ measured at the strongest forcing, $\varepsilon$ = 5. This trace illustrates an extreme wave event ($> 6$ mm wave crest height). The peak amplitude exceeds the standard deviation of the wave background by a factor of more than five, as indicated by a horizontal grey line in Fig.~\ref{fig2}(a). Two movie frames show the waveforms: before the peak, Fig.~\ref{fig2}(b), and during the large event, Fig.~\ref{fig2}(c). The rogue wave is characterized by an almost vertical wave front. Fig.~\ref{fig2}(d) shows the probability density function (PDF) of the normalized wave crest heights $x=\eta_c/\sigma$ ($\sigma$ is the standard deviation) recorded for 300 s, or $10^4$ wave periods. Up to the crest heights of $x =$ 5, the PDF is approximately exponential, $\sim e^{-2x}$. However the strongest waves, $x >$ 5, have a probability which is substantially higher than expected from the $e^{-2x}$ trend. 

An important question related to the high probability of large events seen in Fig.~\ref{fig2}(d), is whether there is a threshold for the occurrence of rogue waves. The fast video technique does not provide sufficient spatial resolution to resolve wave heights at low forcing. To characterize the onset of rogue waves, a more sensitive technique is used, based on the measurement of the intensity of light transmitted through a layer of diffusing liquid, whereby the intensity is proportional to the surface height \cite{Wright1996,Henry2000}. A thin ($<$0.2 mm diameter) laser beam is launched into the container from below, normal to the free surface of the water, as shown in Fig.~\ref{fig1}(b). The transmitted laser light is collected onto the diffusive screen, which is imaged into a photomultiplier tube. This technique is very sensitive to small surface perturbations and it complements fast video imaging required at large wave amplitudes. 

\begin{figure}
\includegraphics[width=6.5 cm]{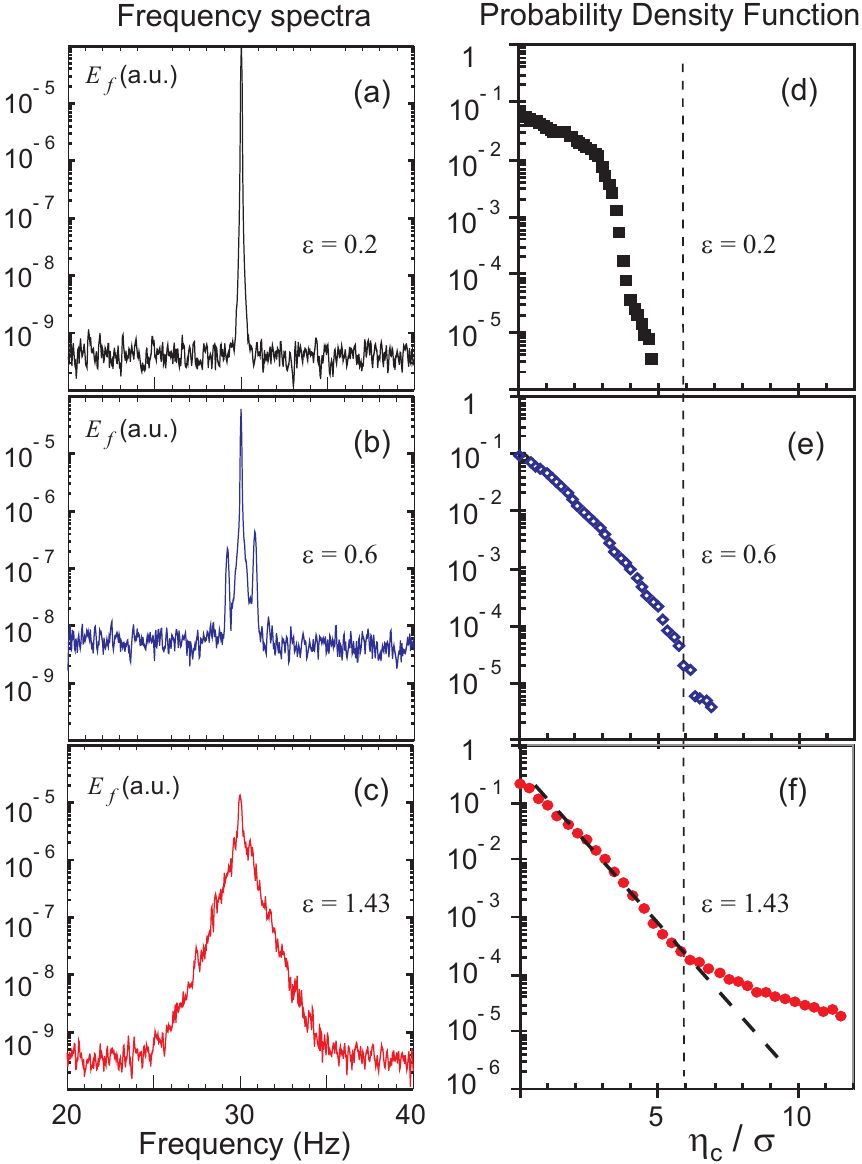}
\caption{\label{fig3} Spectra and probability distribution functions above the threshold of the parametric wave excitation. (a-c) Frequency spectra around 30 Hz and corresponding (d-f) probability density functions at four different levels of supercriticality $\varepsilon$.}
\end{figure}

Figure~\ref{fig3} shows spectra of parametrically excited waves at 30 Hz along with the corresponding PDFs for three levels of forcing starting at $\varepsilon$ = 0.2, Fig.~\ref{fig3}(a). As the forcing is gradually increased we first observe the development of discrete frequency sidebands, Fig.~\ref{fig3}(b), in agreement with classical phenomenology of modulation instability (see e.g. \cite{Zakharov2009}). At $\varepsilon$ = 1.43, Fig.~\ref{fig3}(c), the spectrum becomes continuously broadened showing exponential tails. Further increase in forcing leads to further broadening. The shape of these spectra ("triangular" when plotted in log-linear scale) can be approximated by the hyperbolic secant function \cite{Punzmann2009}. 

A new observation here is that the broadened spectra showing exponential tails occur above a certain forcing threshold ($\varepsilon >$ 1.2). Even more remarkable is the observation of the strongly increased probability of the large wave events, as seen in Fig.~\ref{fig3}(f). The onset of the tails in the PDF of the waves with the crest heights in excess of 6$\sigma$ coincides with the formation of exponential spectra. This high probability of rogue waves is then sustained up to the highest levels of forcing, up to $\varepsilon$ = 5, as in Fig.~\ref{fig2}(d).

\begin{figure}
\includegraphics[width=6.5 cm]{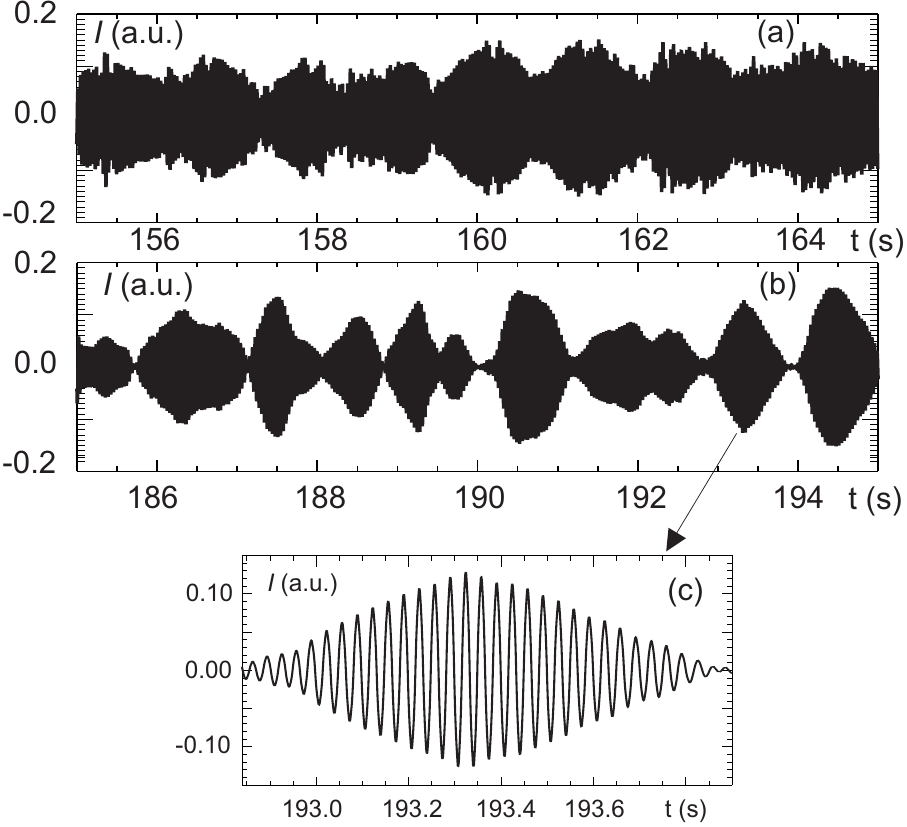}
\caption{\label{fig4} Temporal evolution of the wave height at (a) $\varepsilon$ = 0.6, and (b) $\varepsilon$ = 1.43. The inset (c) shows zoom into one of the envelope solitons.}
\end{figure}

In the time domain the onset of discrete frequency sidebands corresponds to a relatively smooth wave modulation, Fig.~\ref{fig4}(a). Above the threshold of $\varepsilon >$ 1.2 much deeper modulation seen as the ensemble of individual wave envelopes is observed, Fig.~\ref{fig4}(b). This transition occurs above the same threshold as the transition to exponential spectra. Indeed, the inverse Fourier transform of the hyperbolic secant spectra $E_f \propto sech [b(f-f_0)]$ is given by $s(t) = (\pi/b)sech [\pi^2/(bt)]e^{if_0t}$, which is a well known solution of the nonlinear Schr\"{o}dinger equation, describing envelope solitons (see e.g. \cite{Dias1999,Zakharov2004} and references therein). Fig.~\ref{fig4}(c) shows a zoomed-in envelope soliton. Similar envelope solitons are also found in surface gravity waves \cite{Yuen&Lake1975,Yuen&Lake1982}. 

The origin of the exponential spectra has been discussed from the point of view of envelope soliton statistics \cite{Punzmann2009}. Stronger forcing leads to the generation of shorter envelope solitons and broader frequency spectra. The generation of nonlinearly broadened spectra in surface wave experiments resembles spectral broadening of the light pulses in photonic crystal fibers studied in the context of the supercontinuum generation \cite{Dudley2006} at its initial stage. The formation of the exponential spectra in optical fibers is attributed to either the generation of solitons \cite{Kutz2005,Korneev2008}, or more recently, to the nonlinear dynamics of the Akhmediev breathers \cite{Dudley2009}. Our results, in particular the transition from discrete sideband spectra to exponential  spectra, Fig.~\ref{fig3}(b,c), indicate that the change in the spectra corresponds to a transition from smooth quasi-periodic modulation of waves to the formation of ensembles of envelope solitons, as shown in Fig.~\ref{fig4}(c). The rogue wave generation probably results from a process, similar to the collision of breathers \cite{Akhmediev2009}, or it is due to the envelope soliton interactions observed in nonlinear numerical models \cite{Clamond2006}.

\begin{figure}
\includegraphics[width=4.0 cm]{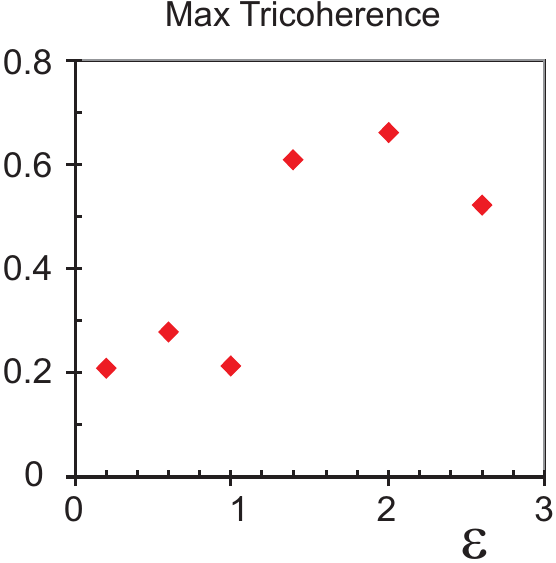}
\caption{\label{fig5} Maximum tricoherence as a function of supercriticality $\varepsilon$.}
\end{figure}

In any case, modulation instability appears as a characteristic feature for such a broadening. Since four-wave interactions of the carrier wave with its sidebands lie in the heart of modulation instability, one needs to detect these interactions to positively identify the instability in experiments. Modulation instability is the process in which sidebands interact with the strong carrier wave simultaneously satisfying the matching rules for the wave numbers, $k_1+k_2=k_3+k_4$, and for frequencies, $\omega_1+\omega_2=\omega_3+\omega_4$. For example, in the spectrum of Fig.~\ref{fig3}(b) one can see three peaks corresponding to the wave, $f_0=30$ Hz, and its two sidebands, $f_{1,2} = 30 \pm 0.9$ Hz. The degree of the four-wave coupling can be characterized by tricoherence \cite{Kravtchenko1995}, or the normalized trispectrum, defined as
 
\begin{equation}\label{Eq:Tric}
t^2(\omega_1,\omega_2,\omega_3)=\frac{\left|\left\langle{F_1F_2F_3F^{*}_{1+2-3}}\right\rangle \right|^2}{\left\langle\left|{F_1F_2F_3}\right|^2\right\rangle \left\langle\left|{F_{1+2-3}}\right|^2\right\rangle},
\end{equation}
\noindent where $F_i$ is the Fourier component of the surface elevation $\eta(t)$ at the frequency $\omega_i$ and $F^{*}_{1+2-3}$ is its complex conjugate at the frequency $\omega_1+\omega_2-\omega_3$. If tricoherence is zero, it is indicative of no coherent phase coupling between the wave quartets, while $t^2$ = 1 corresponds to coherent phase coupling. Fig.~\ref{fig5} shows the maximum value of tricoherence as a function of forcing computed for the transmitted light intensity signals whose spectra are shown in Fig.~\ref{fig3}. The level of the tricoherence is high, $t^2 >$ 0.5, at the forcing above $\varepsilon =$ 1.0 indicating strong phase coupling in four-wave interactions. The level of tricoherence computed for the test noise signal is less than $t^2 <$ 0.05.  Lower tricoherence at lower forcing, $\varepsilon \le 1.0$, may be due to the insufficient frequency resolution in the tricoherence computation ($\Delta f =$ 1.6 Hz), such that the sidebands at $f_{1,2} = 30 \pm 0.9$ Hz were not resolved. The high level of tricoherence confirms the significance of the underlying four-wave interactions. 

It should be noted that the dispersion relation of capillary waves allows simultaneous satisfaction of the three-wave matching rules. This is why three-wave interactions are usually considered the most likely mechanism of the wave-wave interactions among capillary waves. The new and perhaps unexpected finding in our work is that the dominant nonlinearity here is due to modulation instability. High level of tricoherence indicates that 4-wave resonances are at work here and modulation instability is indeed present. Further evidence in support of modulation instability will be reported elsewhere.

Summarizing, we have shown that extreme wave events occur in parametrically excited capillary waves. These rogue waves are very steep. Their heights exceed the standard deviation of the surface elevation by a factor of five to twelve. Rogue waves are represented in the probability density functions as a distinct tail. The rogue wave probability is one to two orders of magnitude higher than expected from the PDF of the wave background. High level of phase coupling in four-wave interactions supports the hypothesis that modulation instability is the key ingredient in the rogue wave generation. This instability leads to a qualitative change in the frequency spectrum of the waves. The onset of exponential wave spectra is correlated with the observation of tails in the PDF. In the time domain the observation of rogue waves is correlated with breaking of the modulated wave into ensembles of envelope solitons. These findings may also help understanding the rogue wave probability and mechanisms of their generation in other wave systems.

\begin{acknowledgments}
The authors are grateful to N. Akhmediev for useful discussions. This work was supported by the Australian Research Council's Discovery Projects funding scheme (DP0881571). 
\end{acknowledgments}

\end{document}